\documentclass[twocolumn,showpacs,preprintnumbers,fleqn,amsmath,amssymb,aps]{revtex4}
\usepackage{graphicx}
\usepackage{dcolumn}
\usepackage{bm}
\usepackage{amsmath}
\usepackage{amssymb}
\usepackage{graphics}

\begin{document}
\title{Uniaxial magnetocrystalline anisotropy in ${\rm CaRuO_3}$}
\author{Moty Schultz}
\author{Lior Klein}
\affiliation{Department of Physics, Bar Ilan University, Ramat Gan
52900, Israel}
\author{J. W. Reiner}
\altaffiliation{Present address: Department of Applied Physics,
Yale University, New Haven, Connecticut 06520-8284}
\author{M. R. Beasley}
\affiliation{Department of Applied Physics, Stanford University,
Stanford}

\date{\today}


\begin{abstract}

${\rm CaRuO_3}$ is a paramagnetic metal and since its low
temperature resistivity is described by $\rho=\rho_0+AT^\gamma $
with $\gamma \sim 1.5$, it is also considered a non-Fermi liquid
(NFL) metal. We have performed extensive  magnetoresistance and
Hall effect measurements of untwinned epitaxial films of ${\rm
CaRuO_3}$. These measurements reveal that ${\rm CaRuO_3}$ exhibits
uniaxial magnetocrystalline anisotropy. In addition, the
low-temperature NFL behavior is most effectively suppressed when a
magnetic field is applied along the easy axis, suggesting that
critical spin fluctuations, possibly due to proximity of a quantum
critical phase transition, are related to the NFL behavior.

\end{abstract}

\maketitle

\section{Introduction}

${\rm CaRuO_3}$ is a paramagnetic metal that attracts considerable
attention due to its intriguing properties; in particular, its low
temperature resistivity described by $\rho=\rho_0+AT^\gamma $ with
$\gamma \sim 1.5$ \cite{CRONFL} and its non-Drude optical
conductivity \cite{CROopt}. These properties suggest that ${\rm
CaRuO_3}$ is a non-Fermi liquid (NFL) metal and  since proximity
to a magnetic quantum critical point could be the source of this
behavior, it is important to elucidate the magnetic properties of
${\rm CaRuO_3}$.

Here we present measurements of untwinned ${\rm CaRuO_3}$ films
and clearly demonstrate for the first time that ${\rm CaRuO_3}$
exhibits anisotropic paramagnetic susceptibility that could be
described in terms of an anisotropic susceptibility tensor with
the easy axis of magnetization at 45 degrees out of the film
plane. This kind of anisotropy is also exhibited by a related
compound, the itinerant ferromagnet ${\rm SrRuO_3}$, in its
paramagnetic phase \cite{SROanisotropy}.

Direct magnetic measurements of magnetic properties of
paramagnetic films suffer from  signal weakness combined with
large background signal of  the substrates. Therefore, we use
indirect magnetic measurements where the magnetic properties are
inferred from measurements of magnetoresistance (MR) and
extraordinary Hall effect (EHE).

Following the identification of the magnetic properties of ${\rm
CaRuO_3}$ films, we found that suppression of the low-temperature
non-Fermi liquid behavior by external magnetic field is most
efficient when the field is applied along the easy axis of
magnetization; suggesting that spin fluctuations are related to
the NFL behavior.

\begin{figure}
\begin{center}
\includegraphics [scale=0.6]{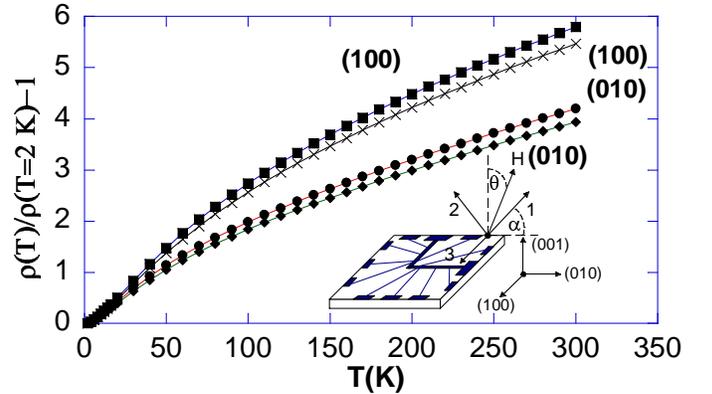}
\end{center}
\caption{Rescaled resistivity of four patterns as a function of
temperature. In two patterns the current is along [100] while in
the other two the current is along (010). Inset: A sketch of the
patterned film. The crystallographic directions 1 (easy axis), and
2 (hard axis) are at $45^0$ out of the plane of the film.}
\label{RvsT}
\end{figure}

Our samples are patterned thin films grown either on ${\rm
SrTiO_3}$ or on ${\rm NaGaO_3}$. The measurements presented here
are  of a 100 nm - thick film of ${\rm CaRuO_3}$ grown on slightly
miscut $(\sim0.3^o)$ ${\rm NaGaO_3}$ substrate with
 resistivity ratio of $\sim6.5$. Resistivity measurements of four
different patterns on the same film reveal correlation between
resistivity curves and current direction (Figure $\ref{RvsT}$).
This indicates that in the films grown on miscut substrates the
intrinsic anisotropy is not averaged by twinning. Similar methods,
enabled growing of untwinned ${\rm SrRuO_3}$ films. For
convenience, we note in the following the miscut direction as the
(010) direction of the substarte.

\section{Measurements}

\begin{figure}
\begin{center}
\includegraphics [scale=0.5]{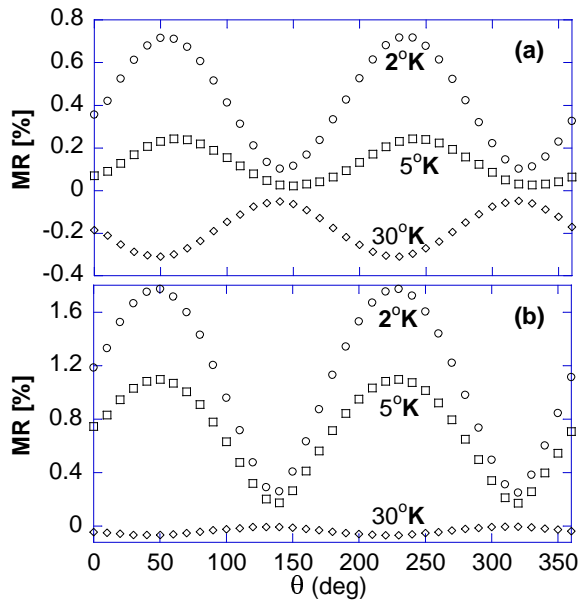}
\end{center}
\caption{MR data at 2 K, 5 K and 30 K for $\mathbf{H}$=3 T as a
function of the angle $\theta$ between $\mathbf{H}$ and the normal
to the sample in the (100) plane. (a) Pattern with current path
along the (010) direction. (b) Pattern with current path along the
(100) direction.} \label{MRvsAnglepatBCandGHf}
\end{figure}

\begin{figure} [h!]
\begin{center}
\includegraphics [scale=0.5]{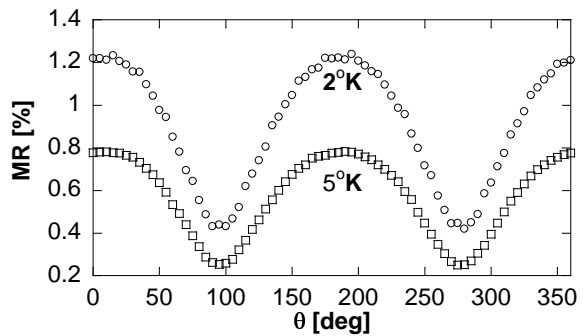}
\end{center}
\caption{MR data for current path along the (100) direction at 2 K
and 5 K for $\mathbf{H}$=3 T as a function of the angle $\theta$
between $\mathbf{H}$ and the normal to the sample in the (010)
plane. } \label{MRvsAnglepatGHs}
\end{figure}

The magnetic anisotropy of ${\rm CaRuO_3}$ is manifested in MR
measurement shown in Figures $\ref{MRvsAnglepatBCandGHf}$ and
$\ref{MRvsAnglepatGHs}$. The MR is measured as a function of
$\theta$ (See Figure \ref{RvsT}) which is the angle between the
applied field and the normal to the film with the axis of rotation
(100) (Figure $\ref{MRvsAnglepatBCandGHf}$) or (010) (Figure
$\ref{MRvsAnglepatGHs}$). We note that at low temperatures the MR
is positive while at high temperatures the MR is negative.
Although the temperature at which the MR changes its sign and the
values of the MR depend on the current direction and the magnetic
field direction, the patterns with current paths along different
directions exhibit nonetheless the same qualitative angular
dependence: with rotation axis along (100) the graph extrema  are
at $\theta=\frac{\pi}{4}+\frac{n\pi}{2}$ while with rotation axis
along (010) the graph extrema are at $\theta=\frac{n\pi}{2}$. This
indicates that the dominant source of MR is related to the
orientation of the applied field relative to the epitaxial film
irrespective of current direction.

The observed MR cannot be attributed to Lorentz MR. At high
temperatures  the MR is negative while Lorenz MR is positive. At
low temperatures, the MR is positive; however, it is strongly
temperature dependent, which is not expected for Lorenz MR where
the resistivity hardly changes. For example between 5 K to 2 K the
resistivity  changes by less than $5\%$ while the MR changes by
more than $100\%$. Moreover, the estimated mean free path even in
the zero temperature limit ($\thicksim30{\AA}$)
\cite{electronicsrycture} is much smaller than the cyclotron
radius ($\thicksim1\mu m$).

Since Lorenz MR is excluded, we look for MR effects related to the
magnetization. Although anisotropic MR is important in this
material, it cannot be the dominant effect, since the qualitative
angular dependence of the MR is insensitive to the current
direction relative to the field. Therefore, a likely source is MR
which is sensitive to the magnitude of the magnetization.

At temperatures above $\sim10$ K the MR is negative and it can be
attributed to suppression of spin fluctuations. Hence, the angular
dependence of the MR reflects anisotropic susceptibility of ${\rm
CaRuO_3}$ with the easy axis at (011) and the two other  principal
axes at $(0\bar{1}1)$ and $(100)$. This explains the observation
that for rotation around $(100)$ the graph extrema are at
$\theta=\frac{\pi}{4}+\frac{n\pi}{2}$ while rotation around
$(010)$ yields graph extrema at $\frac{n\pi}{2}$.

At low temperatures the MR is positive (yet not related to Lorenz
MR) and the angle of maximum negative MR at $T>10$ K turns into
the angle of maximum positive MR at $T<10$ K. Since the source of
the positive MR is unclear at this stage we need other
measurements to determine if the maximum of positive MR also
coincides with maximum magnetization. Namely, to determine whether
the easy axis at $T>10$ K is also the easy axis at $T<10$ K. We
thus turn to measurements of extraordinary Hall effect.

\begin{figure}
\begin{center}
\includegraphics [scale=0.5]{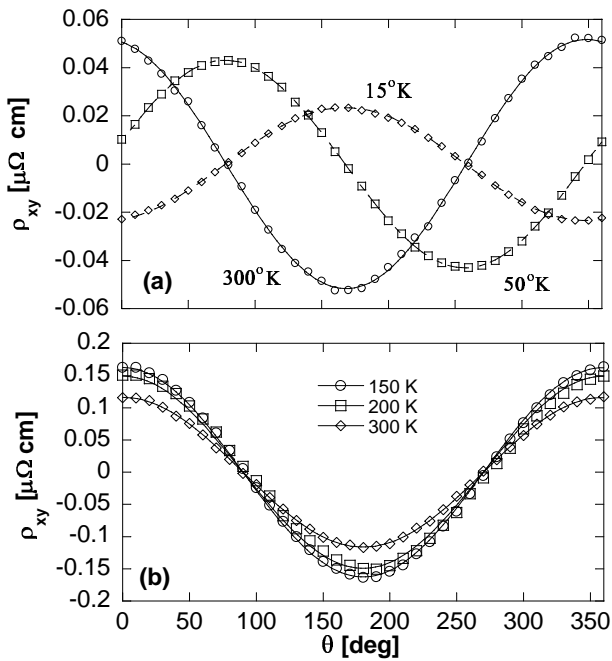}
\end{center}
\caption{HE data as a function of the angle $\theta$ between
$\mathbf{H}$ and the normal to the sample in the (100) plane at
$\mathbf{H}=$ 4 T with curves that are fit of
$\cos(\theta+\varphi)$ (a) and in the (010) plane at $\mathbf{H}=$
8 T with curves that are fit of $\cos\theta$ (b). }
\label{HEvsAnglefs}
\end{figure}

The Hall field in magnetic conductors, ferromagnetic and
paramagnetic, has two contributions:
\begin{equation}
\\\mathbf{E}_H=-R_0\mathbf{J}\times\mathbf{B}-R_S\mu_0\mathbf{J}\times\mathbf{M}
\label{par}
\end{equation}
where \textbf{B} is the magnetic field, and $R_0$ and $R_S$ are
the ordinary and the extraordinary Hall coefficient, respectively.
The angular dependence of the ordinary Hall effect (OHE) is simple
and follows the perpendicular component of $\mathbf{B}$. On the
other hand, the EHE depends on the perpendicular component of
$\mathbf{M}$ whose magnitude depends (in the presence of uniaxial
anisotropy) on the angle between the field and the easy axis.
Figures \ref{HEvsAnglefs}a and \ref{HEvsAnglefs}b present the
total Hall effect (HE) as a function of $\theta$ for (100) plane
and (010) plane, respectively, for several temperatures. The
ordinary Hall Effect is proportional to $\cos\theta$ while the
measured HE has $\cos(\theta+\varphi)$ dependence (Figure
\ref{HEvsAnglefs}a); hence, OHE alone cannot explain the observed
angular dependence. Similarly, EHE cannot account for the results
without the magnetocrystalline anisotropy.

Quantitatively, the magnetic anisotropy is expressed by an
anisotropic susceptibility tensor, $\tilde{\chi}$ with the
magnetization vector given by
\begin{equation}
\\\ \mu_0\mathbf{M}=\tilde{\chi}\mathbf{H}
\label{par}
\end{equation}
where $\tilde{\chi}$ is the susceptibility tensor with eigenvalues
$\chi_1, \chi_2, \chi_3$, corresponding to a field applied along
the three  principal axes.  Because $(100)$ is in the plane of the
film, our HE measurements are not sensitive to $\chi_3$. However,
MR measurements indicate that $\chi_2\sim\chi_3$.

With this notation the dependence of the total HE on the angle of
the applied field $\theta$ in the $(100)$ plane ($\theta$ measured
from the normal to the plane of the film) is expected to be:
\begin{equation}
\begin{aligned}
\rho_{xy}&=\cos\theta H[R_0+R_S(\chi_1\sin^2\alpha+\chi_2\cos^2\alpha)]+\\
& \sin\theta HR_S[\chi_1-\chi_2]\sin \alpha \cos \alpha =\\
& =\frac{1}{2}H(R_1+R_2)\cos \theta+\frac{1}{2}H(R_1-R_2)\sin
\theta
\end{aligned}
\label{EHE1}
\end{equation}
and in the $(010)$ plane is:
\begin{equation}
\begin{aligned}
\rho_{xy}&=\cos \theta H[R_0+R_S(\chi_1\sin^2\alpha+\chi_2\cos^2\alpha)]=\\
& =\frac{1}{2}H(R_1+R_2)\cos \theta
\end{aligned}
\label{EHE2}
\end{equation}
where $R_1=R_0+R_S\chi_1$, $R_2=R_0+R_S\chi_2$ and $\alpha=45^{o}$
(See Figure \ref{RvsT}). The curves in Figures \ref{HEvsAnglefs}a
and \ref{HEvsAnglefs}b are fits to Equations \ref{EHE1} and
\ref{EHE2}, respectively. As we can see, the fits are
satisfactory: At all temperatures the HE in the $(100)$ plane
behaves as $\cos(\theta+\varphi)$ while in the $(010)$ plane as
$\cos\theta$.

\begin{figure}
\begin{center}
\includegraphics [scale=0.5]{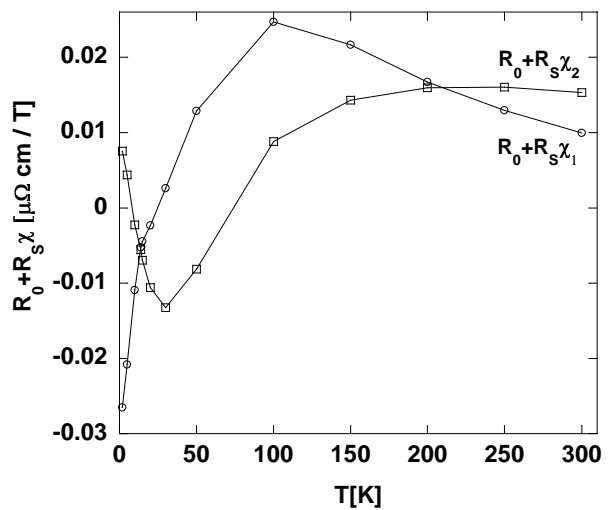}
\end{center}
\caption{$R_0+R_s\chi_1$ and $R_0+R_s\chi_2$ as a function of
temperature} \label{RplusRs}
\end{figure}

As seen in Figure \ref{RplusRs}, $R_1$ and $R_2$ are equal at
$\sim14$ K and $\sim200$ K. There are two possible causes for the
identical values: either the EHE vanishes, ($R_S=0$), or the
anisotropy vanishes, ($\chi_1=\chi_2$). The angular-dependent MR
at $\sim14$ K indicates that $\chi_1\neq\chi_2$; hence  $R_S$
changes its sign at $\sim14$ K. The MR at $\sim200$ K is too small
to measure; therefore, we cannot refute either possibility.
Nevertheless, the fact that magnetic anisotropy is observed at
higher temperatures  suggests vanishing of $R_S$ at 200 K.

If the two crossing points of $R_0+R_S\chi_1$ and $R_0+R_S\chi_2$
are both due to sign change of $R_S$, it means that in the entire
temperature interval that we measure $\chi_1>\chi_2$. This means
that in ${\rm CaRuO_3}$ the easy axis is the $(011)$ direction
which is in our film at angle $\alpha\sim45^o$ relative to the
plane of the film, as illustrated in Figure \ref{RvsT}. The two
other axes are along $(0\overline{1}1)$  and $(100)$ directions.
This means that the maximum positive MR is correlated with maximum
of magnetization. We discuss below possible source of such
behavior.

The temperatures at which $R_S$ vanishes  enable convenient
measurements of $R_0$. Using the most naive one-band calculations
$R_0=(nqc)^{-1}$, where q is the carrier charge and n is the
carrier density, the estimated carrier density is
$n\approx1.2\times10^{23}cm^{-3}$ of electrons at $\sim14^oK$, and
$n\approx3.8\times10^{22}cm^{-3}$ of holes at $\sim200^oK$. Since
the Hall coefficient shows both signs, both electronlike and
holelike bands contribute to the conductivity.

\begin{figure}
\begin{center}
\includegraphics [scale=0.48]{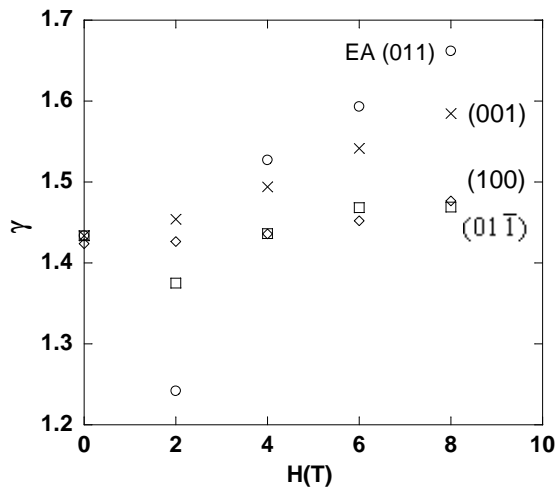}
\end{center}
\caption{$\gamma$ as a function of the magnetic field,
$\mathbf{H}$, applied along three different orientations.}
\label{nvsT}
\end{figure}

Since  proximity to a magnetic quantum critical point could be the
source of the NFL behavior in ${\rm CaRuO_3}$, we explore the
field dependance of $\gamma$. Figure \ref{nvsT} shows $\gamma$ as
a function of field when the field is applied along various
directions. The largest increase of $\gamma$ is obtained for
magnetic fields applied along the easy axis, while the smallest
increase is obtained for magnetic fields applied along the hard
axes. These results indicate that increasing the magnetization is
a route of turning ${\rm CaRuO_3}$ into a Fermi liquid metal.

\begin{figure}
\begin{center}
\includegraphics [scale=0.48]{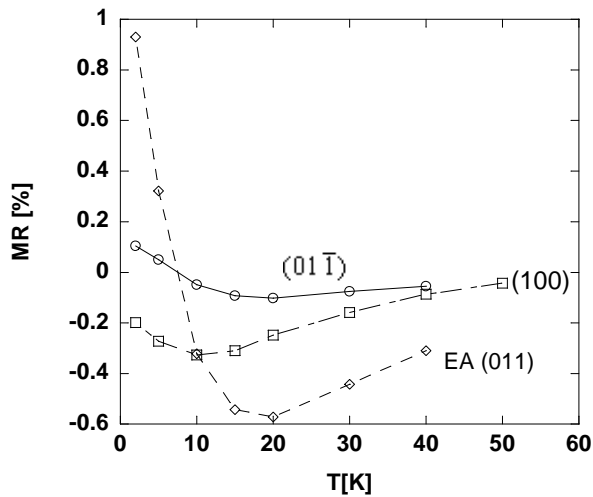}
\end{center}
\caption{MR data as  a function of temperature at
$\mathbf{H}=4T$.} \label{MRvsT}
\end{figure}

Figure \ref{MRvsT} presents the MR with the field applied along
the three principal axes as a function of temperature. As shown in
Figure $\ref{MRvsAnglepatBCandGHf}$ at low temperatures the
maximum of the positive MR is when the field is applied along the
easy axis, $\theta\approx45^o$, and at high temperatures the
maximum of the negative MR is also obtained when the field is
applied along the easy axis. These results imply that the negative
MR, as well as the positive MR, are related to $\mathbf{M}$ and
not to $\mathbf{H}$. It is important to note that the temperature
at which the MR changes its sign depends on the field and current
directions.

\section{DISCUSSION}

As mentioned above, the MR is positive at low temperatures and
negative  at high temperatures, and both, the negative and the
positive MR are related to the magnetization and not to the
magnetic field. In addition, at all temperature range
$\chi_1>\chi_2$. Therefore, the cause for the sign change has to
be due to two mechanisms both relating the magnetization to the MR
effect.

The negative MR is commonly found in magnetic metals, the applied
field suppresses the spin fluctuations, thus yielding negative MR.
${\rm CaRuO_3}$ is a paramagnetic metal. Therefore, one may expect
spin contribution to resistivity and hence negative MR.

On the other hand, magnetization related positive MR is less
understood. We propose that a possible source for this behavior is
the correlation between magnetization and band structure.
Electronic structure calculations of ${\rm CaRuO_3}$ using the
linear muffin-tin orbital  method show a sharp peak in the density
of states (DOS) near Fermi level \cite{electronicsrycture}.
Therefore, a possible source for the positive MR could be the
large variations of the DOS at the Fermi energy. The conductance
in metals is proportional to the density of states at the Fermi
level, therefore applying a magnetic field on ${\rm CaRuO_3}$
which at low temperature is on the edge of spontaneous splitting
of the spin up and spin down bands can change the conductance of
the metal. If the maximum of the DOS is at the fermi energy at
zero magnetic field then applying magnetic field may reduce the
DOS and hence cause positive MR.

The field-induced change in the DOS could be another factor (in
addition to suppressing of critical spin fluctuations expected
near quantum phase transition) in the partial recovery of Fermi
liquid (FL) behavior in ${\rm CaRuO_3}$. In some NFL metals, like
$\rm CeCu_{5.9}Au_{0.1}$ \cite{CeCuAu2}, $\rm CeCuSi_2$ and $\rm
CeNi_2Ge_2$ \cite{CeNiGe1} the FL behavior is recovered at large
magnetic fields. Here at $\mathbf{H}=8$ T although the FL behavior
is not recovered, $\gamma$ is getting closer to the FL value of 2.

\section{acknowledgments}

L.K. acknowledges support by the Israel Science Foundation founded
by the Israel Academy of Science and Humanities.

\end{document}